\documentclass[pra,twocolumn,nofootinbib]{revtex4-2}
\usepackage{color}
\usepackage{amsmath}
\usepackage{amsfonts}
\usepackage{amssymb}

\begin{document}

\title{Quantity and quantitative properties in physics and metrology}
\author{Savely~G.~Karshenboim}
\email{savely.karshenboim@mpq.mpg.de}
\affiliation{Ludwig-Maximilians-Universit{\"a}t, Fakult{\"a}t f\"ur Physik, 80799 M\"unchen, Germany}
\affiliation{Max-Planck-Institut f\"ur Quantenoptik, Garching, 85748, Germany}

\begin{abstract}
I discuss various aspects of the concept of a quantity in physics and metrology and related consideration in reference documents of IUPAP, IUPAC, ISO, IEC, and JCGM.
\end{abstract}

%\today
~~\leftline{CCU-WG-S/2022{\_01\_c}}
\\
\maketitle

A notion of a quantity in physics and metrology is somewhat different. In physics we focus on a quantitative description of Nature, in metrology the focus is on a presentation of the results of measurements. The difference in focussing has terminological consequences.

The quantitative understanding of Nature in physics takes the form of the physical laws, that set relations between different properties.
We use the name `quantity' for any quantitative property, which enters the quantitative laws. For instance, we would consider the electric field $\vec{E}$ as a quantity rather than its components because vector $\vec{E}$ enters various equations as a whole. We use the vector notation because it is advantageous to do so and to consider the vector as a single object and not as a set of several `independent' numbers. To deal with the Maxwell equations we have to differentiate $\vec{E}(\vec{r},t)$. The differentiability is a property of a quantity which is important in physics. Saying that the Maxwell equations relate $\vec{E}(\vec{r},t)$ and $\vec{B}(\vec{r},t)$, we do not mean they relate their values at a location $\vec{r}$ at time $t$. They do not. They relate them as functions of $\vec{r}$ and $t$ or, alternatively, they relate values of their derivatives at $\vec{r}$ and $t$.

We can present $\vec{E}(\vec{r},t)$ in terms of the electric potential $\phi(\vec{r},t)$ and the vector potential $\vec{A}(\vec{r},t)$, which are not measurable in principle. I remind that both potentials enter all the observable equations through their space and time derivatives, that allows for a change in the potentials, which does not affect any observable properties. That is called the gauge invariance and it is an important physical concept. We use such non-measurable objects because it is advantageous to describe natural phenomena in their terms.

The measurability is an important property of a quantity (in physics), but it is not its necessary attribute by default. (One may recall related problems of quantum mechanics.) In physics we would prefer to use term quantity for quantitative properties that cannot be measured, but would specifically emphasize in various context that only measurable quantities may possess `direct' physical meaning.

Often it is said that metrology is a science of measurements, which is not entirely true. Researchers from different areas of science and from physics in particular, are capable to perform their measurements on their own, as they have done for centuries.
What metrology really does is an important work on the presentation of the results of [mostly routine] measurements and on the unification of the language of such presentations mostly for industrial, commercial, and safety purposes and
related legal problems. Roughly speaking, metrology is supposed to provide the naming and the classification of certain existing things suitable for certain purposes. Those things exist by themselves independently on the definitions we choose. `Quantity' is one of terms to describe certain properties which enter quantitative laws of Nature.

In metrology we are focused on the results of measurements and naturally a quantity should allow for their presentation in terms of a product of a numerical value, preferably as a result of an actual or possible measurement, and a unit. The numerical value cannot be differentiated. To differentiate we have to know the related values in a certain small interval. Quantities, defined in such a way, allow for algebraic operations, but not for operations of mathematical [functional] analysis,
that has a simple reason. When a metrologist say `a quantity' he or she rather means the result of a certain measurement of it.
The algebraic equations relating quantity values directly allow for a calculation of the uncertainty through the {\em propagation of uncertainty\/}, which is possible in a straightforward way with neither differential nor integral equations. Besides the question of the units, the uncertainty of the results of measurement is one of the most important problem of metrology and it is reasonable to focuss on such relations that allows one to deal with the uncertainties.

That may be an additional technical problem in terminology since the classification of quantitative properties relies on a possibility for certain algebraic operations on their values. The [normal] quantities have finite values and finite differences of them, while an infinitely small addition is a mathematical abstraction. There are quantitative properties, which are somewhere between ordinal quantities, differential scales, and quantities, which do not allow for algebraic operations with finite quantity values, but allow, e.g., for an addition of an infinitely small value to a finite one. From mathematical point of view, which should be a definitive one for a quantitative description, that is quite a regular situation that, e.g., happens in the analytic geometry.

$\vec{E}$ cannot be presented in terms of a number (and a unit), but needs three numbers (components) and additional references, that describe the directions of the coordinate basis. Actually, different components of a vector, such as the spherical ones, may have different dimensions and to describe them we may need different units. The dimension of a [vector] quantity is the dimension of its length (absolute value) and in general it is not related to the dimension(s) of its components\footnote{Cf. {\sc Note 3\/} to the definition of the {\sc dimension\/} in {\em VIM\/} which states \cite{VIM} that {\em in deriving the dimension of a quantity, no account is taken of its scalar, vector, or tensor character.\/}}. Besides, in special relativity, especially at the basic level, components of a four-vector (i.e., their Cartesian-type coordinates in the Minkowsky space) may have different dimensions, such as $x_\mu=(t,\vec{x})$. Another once-popular presentation makes the 4th component of physical 4-vectors pure imaginary. We definitely are not capable to measure an imaginary number. The result of a measurement is supposed to be a real one and therefore $t$ is a quantity (as well as $c\cdot t$), but $x_4= i\cdot ct$ is rather not. Note, metrological definitions mentioning `numerical value' assume it is real by default (see Appendix). The impedance $Z$ is another example of a widely used complex value which {\em in practice\/} is referred to as a quantity.

Returning to the vectors and tensors, we have to mention that from their sporadic appearances in {\em VIM\/} it is impossible to understand whether the authors mean `our' $3d$ space or $4d$ [Minkowsky] space-time, or an arbitrary one. In many problems in physics and other areas one has to deal with a parameter space, especially when we are interested to use the {\em state vectors\/}. Some of such spaces are fundamental, such as the phase and configuration spaces in classical analytic, quantum, and statistical mechanics, some are empiric. That may be one where algebraic operations are allowed and well-defined, but not the scalar product or the distance between two points. (As a matter of mathematical definitions, the {\sc vector space\/} is the one where the linear operations are allowed, while the {\sc Euclidean space\/} is a vector space where additionally a scalar product is defined.)

A good illustrative example is the $p-T$ diagram of the phases of a water. The components of a $p-T$ vector are the pressure $p$ and the temperature $T$, which obviously have different dimensions. There is no doubts that when we have a vector of any kind, its components can be considered as quantities. But we need to deal with an $p-T$ pair as a whole. E.g., we should be able to study evolution of the system (through its trajectory in the $p-T$ space) and, in particular, to distinguish whether the water is in its liquid or solid phase at a certain moment of time. Another example of the Euclidean space is the application of complex numbers, e.g., for the impedance. The complex numbers can be present with a $2d$ real vector space with the scalar product defined, i.e., any complex number can be presented as a real $2d$ vector.

The appearance of multicomponent properties as a mathematical concept used in physics is caused by the fact that their application is beneficial. They have been introduced for a reason, because it is advantageous to consider them as a single entity, rather than as a collection of different numbers. We have also developed a mathematical language of using vectors of various kinds. Notation, like $\vec{a}\perp\vec{b}$, $v_\parallel$, or ${\rm rot}\,\vec{B}(\vec{r})$, very efficiently utilizes the vector concept.
I would say that is more of a task-related terminology than of a measurement-related one. We are interested in a location in the $p-T$ plot and we intend to speak in terms of the location. Indeed, the measurers should have a good scope of terminology to discuss how to make their measurements and to explain their results. But the measurements are usually done for a reason, which may be of a more complicated nature than each individual measurement of a component of the multicomponent object. We should be able to discuss the results of the measurements in those more complicated terms since that is often what we really need. I believe that a large part of such discussions (on the results in terms of multivariate task-related terms) still lies in the field of metrology. The current definition of a quantity and related terms denies us an opportunity to discuss a multivariate property as a whole, which should require a somewhat different terminology.

It is commonly said in metrological reference books and textbooks that the [numerical] value, involved in the definition of a quantity, is to be a result of a measurement. That means that $\phi(\vec{r},t)$ and $\vec{A}(\vec{r},t)$ are not quantities in a rigorous metrological sense.

One more example of a difference in approaches in physics and metrology is due to general relativity (GR). In its nonrelativistic and zero-gravity limit we may consider Newtonian mechanics within such a coordinate network that in each point the unit of the length is different and possibly time dependent. That is something which is hardly acceptable for metrology, even from a terminological point of view, however, it is a kind of a standard problem in GR. (Roughly speaking bending a light trajectory at a gravitational field is a consequence of a variation of the speed of light in space, which is a natural measure of velocity of the SI \cite{SI}.---The truth is it is a natural {\em local\/} measure in GR.\footnote{I would say that the standard metrology (say, such as presented in {\em VIM\/} \cite{VIM} and the SI brochure \cite{SI} and maintained by BIPM and National metrological institutes) is well prepared for local measurements, but not for the global ones such as navigational at large distances. Instead of the local standards we need for them the time-and-coordinate networks. The job is done by other international bodies such as the International Astronomy Union, IAU.})
Speaking about conceptual differences, the unit of the distance changing in time produces measurable effects in GR through a nontrivial behavior of the affine connection (also known as the Christoffel symbols), which affects the equations of motion etc. On top of that, within GR one can distinguish whether the possible nontrivial behavior in a certain coordinate network is caused by a `real' gravitation or by a `bad' choice of the local units changing in time and space. Within the metrological concept of the units, where only discrete measurements are assumed, it is hard to understand how could one detect a time variation of a unit [of the length].

There are many other differences. Roughly speaking a quantity in physics is what we have in our equations, whatever it is. (That is because in science and in education we do need an umbrella term for all the entries of the equations and we have none other than `a quantity'.) A quantity in metrology is a result of a real or possible measurement, a routine one to a certain extend,
suitable for an efficient consideration of its unit and uncertainty.
Actually both somewhat conflicting approaches are realized in various metrological reference books\footnote{Metrological references, cited here, are not available in the libraries of most of physical institutes, so I give for them the titles and the related on-line references if available.}. The definition of a quantity there mostly follows the `metrological concept' (see \cite{ISO,iupac,iupap}, cf. \cite{SI,VIM,GUM}\footnote{{\em GUM\/} \cite{GUM} in contrast to {\em VIM\/} \cite{VIM} is not a single document but has several {\em Supplements\/}. E.g., {\em Supplement 2\/} \cite{Supp2} gives `A glossary of principal {\em symbols\/}' used in there, which assumes that the related {\em terms\/} have been defined somewhere else or have to be, which to the best of our knowledge of the JCGM documents is not the case. In particular, the symbols are given for a complex quantity and for a vector one, that both have been defined neither in {\em VIM\/} \cite{VIM} nor in {\em GUM\/} \cite{GUM}.}),
while an extensive list of quantities (and their symbols) given in some of those references (see \cite{ISO,iupac,iupap}) follows the `physical concept'. The `metrological concept' is the one that is presented at {\em VIM\/} \cite{VIM}, produced by JCGM\footnote{JCGM is the Joint Committee for Guides in Metrology, working groups of which issue {\em VIM\/} \cite{VIM} and {\em GUM\/} \cite{GUM}. IUPAP is the International Union of Pure and Applied Physics; while IUPAC is the International Union of Pure and Applied Chemistry, both of which produce their reference books such as \cite{iupap,iupac} on symbols and nomenclature. ISO is the International Organization for Standardization and IEC is the International Electrotechnical Commission which publish their {\em Quantities \& Units\/} \cite{ISO}.}. The reference books that claim a {\em VIM\/}-compatible definition, but present a list of quantities, contradicting to it, are published by ISO, IEC, IUPAC, and IUPAP, i.e., by 4 of 8 members of JCGM.

Discussing {\em VIM\/} \cite{VIM}, we remind that {\sc Note 5\/} to the definition of a {\sc quantity\/} states
\begin{itemize}
\item[]`{\em A quantity as defined here [in VIM] is a scalar. However, a vector or a tensor, the components of which are quantities, is also considered to be a quantity.\/}'
\end{itemize}
That does not help. We take such a statement as an acknowledgement that the worlds of the quantities is much larger and more diverse than that of the [`scalar'] ones, which have been the only ones considered  in {\em VIM\/} so far.

The problem is not that a quantity is formally defined there as a single-numbered one, but that such a definition is a cornerstone one and it is subsequently used as a base in development of a system of definitions. That is not about which kind of quantitative properties we call a quantity, that is about for which kind of them we develop the scope of subsequent terms. Scales, ordering, intervals, uncertainties, magnitudes, relative measurements (ratios) are either introduced only for a real single-numbered quantity or their details are critically different for single-numbered ones and the others. (See Appendix for details.) A sporadic mentioning in {\em VIM\/} of vectors and tensors does not clarify the situation but rather produces a certain inconsistency.

%%%

Considering an extended definition of the quantity, applicable to vectors etc., we could arrive at various uncertain terminological situations. E.g., if we use term `quantity' both for the position vector $\vec{r}$ and for the distance $d$, we may wonder whether they are quantities of the same kind. They are definitely quantities of the same dimension, also definitely we can say that $\vec{r}^2$ and $d^2$ are quantities of the same kind, but I am not sure that we have a consensus that vector $\vec{r}$ and scalar $d$ are quantities of the same kind by themselves. Besides, saying, that they are quantities of the same dimension we note that we cannot measure their ratio. Neither we can measure a ratio of two position vectors. That means that consequences of being two quantities of the same dimension for the extended definition of the quantity are not the same as for the standard one (for a real single-numbered quantitative property).

%%%

{\sc Note 3\/} to the same definition \cite{VIM} states
\begin{itemize}
\item[]`{\em Symbols for quantities are given in the ISO 80000 and IEC 80000 series {\sc Quantities and units\/}.\/}',
\end{itemize}
which means that {\em VIM\/} recognizes as a quantity whatever is called `quantity' in the ISO-IEC listing \cite{ISO}, despite they may not fit the {\em VIM\/} own definition. (The ISO-IEC materials \cite{ISO} are commercialized, they are not for a free distribution, in contrast to {\em VIM\/} \cite{VIM}. I believe an open-access normative document, such as {\em VIM\/} \cite{VIM}, should never rely on any commercialized one. Besides, the hierarchy of JCGM and ISO-IEC documents remain unclear. They cite each the other. In particular, {\em VIM\/} \cite{VIM} cites \cite{ISO} through the mentioned Note 3. But what is more important than the cycling in the definitions, is that the ISO-IEC documents have not been formally approved by the other member-organizations of JCGM, which made them rather questionable as a part of JCGM regulation.)

{\sc Example 5\/} to the definition of the {\sc quantity value\/} \cite{VIM} contains a complex value of the impedance (see also {\sc Note 2\/} there), however in many subsequent definitions the value of a quantity without any reservation demonstrates properties that take place only for real numbers (cf. definition of the uncertainty and various intervals---see Appendix).
If the complex values of quantities are permitted by its definition, then for the subsequent description of quantity's properties, that requires the reality of the numerical value, such a condition should be always explicitly told, which is never a case for {\em VIM\/}.

In other words, if the complex values of a quantity are considered as legitimate, then by default it should be expected that the values of any quantity of interest are complex unless it is explicitly stated otherwise.
Commenting Note 3 and Example 5 of the definition of a quantity in {\em VIM\/} \cite{VIM}, let's remind that usually it is considered by default that a [standard] quantity possesses all the properties of an ordinal quantity (as defined in \cite{VIM}), i.e., its values may be ordered. That is not possible for a quantity with complex values. Neither that is possible for a vector quantity. That means that a decision whether a vector or a complex value may be a quantity is not just a standing-alone terminological question. It should have numerous terminological consequences. The relation between quantities and ordinal quantities plays an important role in description of the logics of the classification of quantitative properties. Besides, for vectors we also cannot define a {\em ratio\/} of two values of [vector] quantities, while for a standard quantity it is usually expected that such a ratio is well-defined and often can be directly accessed for a measurement.

Another important property of a real quantity that we usually accept by default is that the value of quantity has a central value and an uncertainty, which is not applicable to a multicomponent quantity (for the uncertainty). The latter either described with an area where the `true' value can be (say, the standard ellipsoid) or can be parameterized by the uncertainties of each component {\em and\/} their correlation. E.g., the value of the complex impedance has a two-component central value, but its uncertainties for the complex description require {\em three\/} parameters, not two. (See the Appendix for detail.) Therefore to `upgrade' several numbers to a `vector', we can easily combine a `vector' for the central values, but for the overall uncertainty ne has to do something else. Making a vector of several numbers as suggested in {\em VIM\/} by telling that a vector, components of which are quantities, is a quantity, is not a mechanical operation. If the vector quantities are allowed the issue about the correlation of the uncertainties of their components should be explicitly mention. The same relates to the complex numerical values.

Since the legal regulation concerns the results of measurements, one may think that the metrological approach is the one required the `official' definitions, while physicists may resolve their problems without any official reference book as they always did.\footnote{The {\em Red book\/} of IUPAP \cite{iupap} is not the most popular reading of the physical community. That is probably why it has not been updated since 1987.} That is a correct view, but not a complete one.

Still, there is one more area where the regulation may be important. That is the school and university education. The purpose of a broad (non-specialized) part of the education is to prepare people for life in society. The metrological regulation is the one for a certain part of that life in areas related to production, sale, and usage of various devices and other goods, safety, etc. That means that the educational concepts should be consistent with the metrological ones. On the other hand, the education should be based on science and in particular on physics and the base education should serve as the base of professional education in various areas and in particular in physics.

Writing one of the Newtonian laws as
\[
\vec{F}=m\vec{a}\,,
\]
the teachers and students should use a correct term for the involved vectors. Either they are quantities or something else.
Writing the same law in term of momentum
\[
\vec{F}(t)=\frac{d \vec{p}(t)}{d t}
\]
they should understand how the definition with a product of a measured number and a unit is related to the derivative. Results of measurements cannot be differentiated. Measurements allow for discrete finite sequences of the results, they can produce neither a continuous function nor a limit. A series of the results can produce an approximation of a derivative in one or other way, but not the derivative itself. The limit as a mathematical operation cannot be realized on results of real measurements. Metrology cannot allow for a consideration of the equation above as a relation between quantities $\vec{F}(t)$ and $\vec{p}(t)$ at certain $t$,
but as a relation between $\vec{F}(t)$ and a derivative of $\vec{p}(t)$ at $t$. Meanwhile, in physics, we see the equation as a relation of two quantities $\vec{p}(t)$ and $\vec{F}(t)$ as two functions of $t$, and that is what is taught at school and university.

Metrological documents tell many things about quantities, dimensions, units, etc. If some physical objects, such as the radius-vector $\vec{r}$, the [complex] impedance $Z$, or the electric potential $\phi({\vec r}, t)$, are not quantities (in a rigorous metrological sense), teachers and students would be confused and have a problem to understand to what extend various statements from metrological documents are applicable to them. In particular, dealing with an unmeasurable quantitative property, such as an electric potential or a complex value of the impedance, what may we say about their units and uncertainty?

As a matter of fact, physical relations are in general rather not expected to be measured in the common sense. The aim of a quantitative description in physics is not to describe certain phenomena in time and space in a measurable way, but to produce measurable predictions. E.g., one may develop a theory of the interior of Sun while studying properties only of its surface. In its turn, a successful theory of the interior would allow for predictions of changes the temperature of the surface.

A calculation with certain quantities, that mostly cannot be measured or are not supposed to, as a way to reach at some stage certain measurable values is completely ignored in the core metrological definitions. The quantities that appear in an intermediate stage of calculations are rather quantities in a broad physical sense, while the eventual measured results deal with the values of the quantities that sound consistently with the standard metrological consideration. (More than that, definitions of {\em VIM\/} \cite{VIM} relate to quantities and values that can be measured by classical means and ignore the key questions of quantum measurements etc., while the calculations for values that may be obtained by means of classical physics for the related readout are often based on a heavily quantum theory. The quantum theory of measurements considers a measurement as a certain interaction of a quantum system of interest with a classical device that provides the readout. E.g., one has to deal with the Schr\"odinger equation to obtain even approximate values of the energy levels in the hydrogen atom, while the readout of the frequency measurements is a classical one.)

Hopefully, it is declared that the quantities allow for certain mathematical operations on them, which may eventually produce theoretical results. Most of the theoretical results require an experimental input, however, that does not turn a complicated theoretical prediction into the result of a measurement. If we would consider all theoretical results as a result of measurements because some experimental input is used that would produce a confusion.
Roughly speaking theoretical calculations establish relations between observable properties and a part of such relations can be formulated without any experimental input parameters. For instance, we can formulate the Newtonian law of gravitation in part as a statement that when we double the distance the force decreases fourfold. Some of such predictions are for quantities which can be measured (and therefore the prediction is a prediction of the result of a possible measurement), some not (and the `prediction' is an estimation of unmeasurable). The equivalence principle (and the universality of the free fall)
is another example of a quantitative law which does not need any experimental parameter for its prediction. Note, considering theoretical quantities we may easily have a differentiable quantity, which is not possible for [a discrete sequence of] results of measurements.

Concepts of quantities, units, dimensions, and the measurability are very important for physics and physical education. Physicists do not care about formal definitions, but they do care about the contents of their terms. It is a normal situation when a word used as a common word and as a term has different meaning. Quantities, units, dimensions, etc. are not only common words, but also base terms of physics. (Meanwhile, quantities, numbers, values are also base terms in mathematics.) That is a problem that some words for base terms of physics, mathematics, and metrology may have different meaning, but the different versions of the terms (with the same word, say, a quantity) may be in use in the same area [of education or applications] and in the same documents (see, e.g., \cite{ISO,iupac,iupap}) considering physical quantities.

An agreement of physical and metrological approaches in the definition of quantities is not important by itself, as far as two communities work separately. However the educational purposes may strongly constrain the metrological definitions. After all, a number of people who have learned at school about certain vector quantities, such as $\vec{r}$, is highly likely larger than a number of people who need to formally deal with quantities and have never known on $\vec{r}$.
That is why ISO, IEC, IUPAC, and IUPAP \cite{ISO,iupac,iupap} came to the usage of the term of a quantity in their documents in a contradicting way attempting to combine the definitions, designed to provide a formal clarity for metrological applications, and the nomenclature to cover practical needs by considering quantitative properties which are in a broad use.

Recently a new round of discussions on a possible definition of the core metrological terms started within the metrological community. Because of that there are several practical questions to the international organizations which are responsible to produce the final `official' definitions.

(i) {\em Do we really want such terms, as quantity, unit, dimension, to be understood differently in metrology and in physical science and education?\/} The concepts used in physics will not change. Those concepts are in use and need terms. If it would be managed to use the metrological terminology (say, a metrological definition of a quantity) at school that would help not much by itself, because we should still be in need for an additional general term or terms for objects like $\phi(\vec{r},t)$, $Z$, and $\vec{r}$.

(ii) It is reasonable to expect that the [base] physical education will follow the physical science, which leads to another formulation of the question above. {\em Do we really want the definitions in the SI brochure
to be partly inconsistent with the related concepts used in the [base] physical education?\/}  We have in part already encountered such a situation. The concept of the {\em base\/} [physical] quantities is a very useful one for dimensional analysis and education. The base quantities are introduced as the base quantities in description of natural phenomena. (We remind that the description of physical phenomena often includes various derivatives and not all of them have names and recognized as separate physical quantities such as the velocity and the acceleration.) They may be not really needed for the {\em SI brochure\/}. This term has existed not because of the base units and was introduced not through a metrological regulation. The latter has just acknowledged its existence
and attempted to make a use of it for itself. The question is not about the existence and use of the term, but only whether it will be a part of the {\em SI brochure\/} or not.

(iii) The {\em SI brochure\/} may have a certain use at school and universities, but rather indirectly. (If it were of a direct use there, it should be one of the most read books in the world. It does not look like that.)
{\em If we choose to use its definitions more close to the one from the physical science and education, what is to happen with other metrological documents, such as VIM?\/}

After all, the SI is not a system of units, but a system of units of [physical and not only] quantities. That means a clear logical line with the quantities, the metrological community would like to regulate. The quantities, whatever they are, should be specified at the first step, while their units are to be introduced at the next step as a part of such a regulation. The definition of a quantity is very much a strategic choice on what kinds of scientific phenomena and practical activity the metrological community is to cover and what they are to put aside. Unless a certain kind of quantitative properties is classified and named it cannot be a subject of any regulation. To regulate something we have to name it first.

(There is another question, that is in part the same, but only in part. Different mathematical tools have been introduced by mathematicians and applied in physics because their use is beneficial. They create various frameworks, that are a kind of languages, such as the mathematical analysis of functions of a {\em complex\/} variable, the {\em vector\/} algebra, or the Riemann geometry. {\em Depending on the metrological definitions, some of these languages may appear to be inconsistent with metrology, e.g., since their entries are not quantities.\/}---In contrast the discussion of the quantitative properties, here is a consideration of their description.)\footnote{\color{red} One should clearly understand the relations between the terminology and reality. The latter does exist by itself. Scientist do what they do. And they have certain concepts on what they are doing. Terminology may name those concepts and recommend a certain regulation for the use of the terms. The only question is whether a terminological document and the regulators body, that issued it, could regulate certain activity in the real world.}

(iv) {\em Quantity calculus\/} plays an important role in metrology and in particular in metrological documents, and the {\em SI brochure\/} is a key guide for it. It can be efficient if the most of the key relations of interest [of metrological community], that include quantities, would include only quantities. If some of them include quantitative objects other than quantities then we can use those relations to determine a quantity from non-quantities, which would undermine the very idea of the well-controlled system of {\em quantity calculus\/}. The quantities should create a more closed logical structure. After all, multivariate quantitative properties do exist and they are in use in physics. In particular, we can use them to express results of measurements, which means a possibility to satisfactory describe the results in terms that are not covered by metrological regulation at all.

Metrological activity is hardly possible without a deep involvement of fundamental [physical] theory. E.g., to recently fix the values of $e, h, N_A$, and $k$ \cite{si2017,codata2017} for the brand new definitions of the SI a number of complicated experiments were performed and some of them required a heavy theoretical support. Theory is based on physical laws, i.e., on various relations between the quantitative objects which often cannot be measured directly and sometimes cannot be presented as a product of a number and a unit.

In the new SI \cite{SI} the magnetic constant of vacuum $\mu_0$ plays an important role. In contrast to the previous SI version its value is not adopted exactly by the definition, but has to be found from experiment. The value cannot be measured directly with a sufficient accuracy. However, it can be determined from experiments if we use quantum-electrodynamics theory either of the energy levels of the hydrogen atom or of the anomalous magnetic moment of the electron \cite{codata2018}. As one of
quantum-electrodynamics experts, I should confess that I do not understand how to perform any required calculations without using the Green's functions of an electron and of a photon and how to measure those functions. The present-day version of the {\em SI brochure\/} \cite{SI} does not help to understand what are the dimensions of the involved `objects', since they are not quantities in the rigorous sense as defined in \cite{SI}. That sets an interesting example of an important metrological result obtained
within a formalism which is not covered by the {\em SI brochure\/} \cite{SI} and other metrological regulations.

The basic-terms part of {\em VIM\/} is out of interest of both the physical community and a big part of the metrological one, which is presented by the national metrological institutes (NMI). (It is not a secret that the practical metrologists from NMI's are mostly not in favor of that part of {\em VIM\/} \cite{VIM}.) The definitions of {\em VIM\/} are not in conflict with the physical science or education, because they have never met each the other. But the {\em SI brochure\/} and {\em VIM\/} may, in principle, collide because both pretend to be key metrological documents, which in a perfect case should present a coherent set.

Concluding, I would like to stress that there are two completely different problems on the definition of a quantity. One is on the concept which may be explained and discussed in loosely defined terms. There are a number of simple questions, whether, say, $\phi(\vec{r},t)$, $Z$, or $\vec{r}$ belong to quantities or not. And if they do, do they belong to the same class of quantities as, say, the temperature and mass? If they do not, what are they? Without a consensus on such questions it would be hard to proceed.

The other problem is an appropriate verbal formulation of the definition once the conceptual part is understood and agreed upon. The definitions should leave no room for questions such as whether `number' in a definition is a real one, or may also have a complex value. Or whether the `magnitude' is the value or the absolute value? All that should be clarified within the definitions.

It is also crucially important for finalizing the conceptual part to make a decision, whether the {\em SI brochure\/} and metrological definitions given there, are aiming their consistency with physical education at school and university level.
Physicists and metrologists have focused on different aspects of the quantitative properties and their approaches do not coincide.
The guidelines for the physical education (in its part related to the units and uncertainties) is not a part of the responsibility of the metrological community and in particular of CIPM and JCGM, but it would be reasonable to have a certain level of consistency between the physical and metrological approaches to the terminology.

I do not think we have a perfect understanding and agreement on the concept of the quantity within the metrological community.
Let's remind that ISO, IEC, IUPAC, and IUPAP have no consensus, or rather no integrity on the issue inside their organizations since their own documents are controversial on the definition and the usage of term `quantity'. The mentioned organizations recognize their role not only in the definition of the core metrological terms but also in recommending the nomenclature and terminology related to broad science and education areas (including `real' science and professional scientific education), that involve even more complicated constructions (like GR tensors, wave functions, and vectors and operators in Hilbert space) than required for the base education.

In order to resolve the controversy inside those reference documents, the issuing organizations have to give a term (or several terms) for multivariate quantitative properties and unmeasurable quantitative properties (if those are not just quantities) and to explain which part of statements about the quantities of the {\em SI brochure\/} \cite{SI}, {\em VIM\/} \cite{VIM}, and {\em GUM\/} \cite{GUM}, that they mostly consider as their source of definitions, is valid for them.

Consensus within JCGM and CCU\footnote{CCU is the Consultative Committee for Units of the International Committee for Weights and Measures (CIPM). BIPM (the International Bureau of Weights and Measures) publishes the SI brochure \cite{SI} on behalf of CIPM.}, which is required to proceed with {\em VIM\/} and the {\em SI brochure\/}, respectively, depends on the position of four mentioned organizations, which in an ideal case should be based on their integrity and consensus. Otherwise, we may have a consensus within various horizontal [inter-organization] commissions, but an inconsistency between positions of those commissions and in some documents of the member-organizations.

Once it is said that a quantity can be presented as a product of a dimensionless quantitative characteristic and a unit, the terminology is built to cover two crucial cases, namely, the consideration of the units for
a certain broad kind of quantitative properties and the consideration of the accuracy of the determination of the numerical characteristics, which in metrological practice are mostly single-numbered ones.
We have to find a certain balance between the positions of organizations (and communities), which are mostly interested in terminology on the uncertainty of routine measurements, and of the ones, which are interested in terminology and nomenclature on general physical terms for education and science and in particular on the units and dimensions of quantitative properties.

Probably the most suitable solution would be to separate real single-numbered quantities and multivariate ones. That would allow for the usage of the most of the logical construction of {\em VIM\/} for the real single-numbered quantities. As concerning the multivariate ones we have several options. We can completely remove any mentioning of them from {\em VIM\/} and ignore them furthermore, which would make {\em VIM\/} selfconsistent, but not supportive for the base education and not consistent with terminology in physics. Alternatively, we can define both real single-numbered quantities and multivariate ones in such a way that we could use the definitions to define the dimensions and the units in a way suitable for education and physics, while the part of {\em VIM\/} related to the uncertainty etc. is to be considered as the one for the real single-numbered quantities. Such a solution would allow us to prepare the definitions related to the accuracy and the uncertainty of multivariate quantitative properties some time later, but for the moment will leave the logical structure of {\em VIM\/} and its relation to {\em GUM\/} mostly intact. (Technically, {\em GUM} \cite{GUM} does not define basic terms but introduce the symbols for them used in there. Using appropriate symbols would allow one to formulate certain statement only for a certain group of quantities.)

It may be worth to summarize the problem with a practical example considered in simple words, additionally to more or less formal statement.
%To finally conclude the paper. let's consider consequences for an example.

When we consider a simple relation like
\[
v_x=\frac{dx}{dt}
\]
depending on the definition of term `quantity' we can say that the identity sets a relation between quantities $x$ and $v_x$ (if the quantity can be considered as a continuous function of [quantity] $t$ and therefore the differentiation on the quantity is possible) or cannot (if we consider a quantity as a kind of set of possible values and the relations between the quantities are reduced to the relations between the values of two quantities)\footnote{Here, I mostly discuss quantities in physics and metrology. Physics mostly speaks in mathematical terms. In mathematics, the differentiation and integration on [mathematical] quantities are legitimate operations and they are the key ones for certain areas of mathematics.}. Actually, that creates a dilemma. If the equation above sets a relation between the velocity and the coordinate and they {\em are\/} quantities, then a relation between two quantities does not necessary require a relation between their values. That would contradict to the standard consideration, that the relation between two quantities means simultaneously two relations, one between their values and the other between their units, that is its turn is a key for the consideration of the units for related quantities. If we maintain the latter, we should either deny the relation above as the one between quantities $x$ and $v_x$ or should acknowledge that at least one of the entries of the identity is {\em not\/} a quantity. In this pair of equations, one on the values and the other on the units, the first one is vulnerable, because to find a value of $v_x(t_0)$ we need to deal with values of $x(t)$ in a certain interval of $t$ (i.e. to related {\em a value\/} of $x(t_0)$ to {\em values\/} of $v(t)$ in a certain range), however, the equation on units looks intact. The latter {\em looks\/} so, but not necessary is. We can present a value of a quantity as a product of a numerical value and a reference as explained in numerous metrological documents. There is no problem for the reference to be time dependent. That happens all the time. If we need to find a derivative we have to differentiate both the numerical value and the reference, which would mean a somewhat extended approach to `quantity'.

If we made a bad choice and use a time-dependent reference as a unit we have also to introduce its derivative. (We {\em do\/} know that some natural units, such as the period of rotation of Earth around Sun (i.e., a year) or around its own axis (i.e., a day) are time dependent.---Somehow, it is `normal' to expect that a reference in a specific measurement is time dependent, but it is not `normal' to think so about a {\em unit\/}, that from a practical point of view is just a reference with all its possible problems.) Returning to the applications, the first law of Newton speaks about a motion with the constant velocity. That does not mean the same numerical value if the unit of velocity is time dependent. (See textbooks on GR for detail.) Actually, the metrological approach to the units is that we are interested in local measurements, so the units are localized in time and space. That is why to deal with derivatives would create a number of metrological problems for which the terminology is not well developed.

Besides, we have an additional problem with the dimensions. What is $d/dt$? It is an {\em operator\/}, but it is a {\em dimensional\/} operator. We can discuss its dimension, the related unit etc., but we cannot separate the unit and the value of the operator. (In the best case scenario, we can consider such an operator as a product of a dimensionless operator and a unit.) In other words, the dimensional objects exceed the world of quantities.

It would be an interesting formal situation if, rigorously following metrological documents, metrologists should say that they do not know in what terms the physical textbooks are written, because a big part of the description is with the massive use of objects that are {\em not\/} [physical] quantities.

\appendix

\section*{Appendix: some examples on the definition and the subsequent usage of the term `quantity' in {\em VIM\/} \cite{VIM} \label{s:vim:ex}}

According to the {\em VIM\/} definition \cite{VIM}, a {\sc quantity\/} is a {\em property of a phenomenon, body, or substance,
where the property has a \underline{magnitude} that can be expressed as a number and a reference\/}.\\
Since the definition does not say that a quantity does not have a direction or so, one may speculate whether it is applicable to quantitative properties which have a magnitude and something else. Besides, the magnitude is rather to be expressed as a real number.

It is obvious that to define the base terms one has to rely on something which is undefined. Such a definition should be clear and the words and concepts used should be commonly accepted. As we see, we may wonder whether `number' in the definition is assumed to be a real one, or a complex number is also allowed.
We also noticed that term `scalar', used in Note 5 to this definition, is not a formally defined metrological term. In physics it has a different (from \cite{VIM}) meaning. The latter, following {\em VIM\/} \cite{VIM}, consider a scalar as a component of a vector. In physics and mathematics a component of a vector is {\em never\/} considered as a scalar. (Scalar in physics is an {\em invariant\/}. A scalar product of two vectors is a scalar as well as the length of a vector, while its component is not.)
Due to that we cannot be 100\% sure what is literarily said in {\em VIM\/} \cite{VIM}, which is bad by itself. We have to interpret their definitions.
(Term `vector' is also used not with the same meaning as often in physics. E.g., a consideration of vectors in descriptive geometry allows for introduction of coordinates but that is an option, not a necessity and in physics we do not hesitate to use the descriptive geometry. A notation like $\overrightarrow{AB}$ clearly illustrates this notion. Vectors in physics and mathematics often exist in various problems without any consideration of their components.)

In the main text of the paper we give examples on an extended understanding of the definition of a {\sc quantity\/}. In particular, we mention Note 5 to the definition of a {\sc quantity\/} which speaks about extending of the definition to vectors and tensors. Literarily it claims that a quantity {\em is\/} defined in {\em VIM\/} \cite{VIM} as a scalar one, but we {\em may\/} extend its definition to vectors and tensors, but it is unclear to what extend.

Since a {\sc quantity\/} {\em is\/} defined as an one the value of which can be presented as a single real number (and a unit), we can see it from a subsequent introduction of the terms that are defined through the base terms.
Several examples from {\em VIM\/} \cite{VIM} which clearly indicate that a quantity is assumed to have its value as a single real number are given below.

The {\sc quantity value\/} is defined as {\em \underline{a number} and reference together expressing magnitude of a quantity\/}.

Note 2 to the definition of the {\sc measurement result\/} states that {\em a measurement result is \underline{generally} expressed as a single measured quantity value and a measurement uncertainty\/}.

If we allow for vector quantities or complex numbers, the general case is a multivariate one with several real numbers. If the uncertainty is defined as at {\sc VIM\/} (see below) as a single `non-negative parameter' that is also not a general case.

In the definition of a {\sc measurement unit\/} it is said that a unit is a {\em real scalar quantity\/} ..., {\em with which any other quantity of the
same kind can be compared to express the \underline{ratio} of the two quantities as a \underline{number}\/}, which is obviously not valid for vector quantities.

The {\sc quantity value\/} is defined as a {\em number and reference together expressing magnitude of a quantity\/}.

That being applied to a vector $\vec{r}=(1,2,1)\,{\rm m}$ in a certain coordinate frame, makes it value as $\sqrt{6}\,{\rm m}$ or so, because it is related to a magnitude. From the mathematical point of view we should rather expect that $(1,2,1)\,{\rm m}$ is the value. Speaking about vectors, to distinguish in common terms vectors from non-vectors it is often said, that a vector has a magnitude and a direction. It such terms it sounds like the quantity value of a vector deals with its magnitude and ignore its direction.

We recall that the {\sc measurement error\/} is defined as a {\em measured quantity value minus a reference quantity value\/}, which should have not much sense for a vector quantity if its value is its magnitude. The {\sc measurement accuracy\/} and {\sc measurement trueness\/} also rely on a comparison of quantity values.

The {\sc measurement uncertainty\/} is defined as a {\em non-negative parameter \underline{characterizing the dispersion} of the quantity values.\/}

In the case of a multivariate quantitative property, such as a would-be `vector quantity', the dispersion is characterized by a matrix, not by a single parameter. A situation with a dispersion of the values of a complex quantity is also not with a single `non-negative parameter'.
Besides, there are mutlivariate properties, such as the position in the $p-T$ phase diagram, where the uncertainty is characterized with uncertainty in $p$ and $T$ and their covariance. The 3 values have different dimensions and it is not possible to characterize simultaneously all of them with a single number.

Alternatively one may say that one measurement gives one number and to measure a vector we need several measurements, but in such a case the value of a multivariate quantitative property can neither be found by a measurement nor have an uncertainty. A measurement and an uncertainty are possible only for a component of a multivariate quantitative property, which I am not sure is intended.

The {\sc coverage interval\/} is defined as an {\em interval containing the set of true quantity values of a measurand.}

That has sense only for quantities, the values of which are single real numbers, otherwise we have to speak about an area of possible values.

\onecolumngrid

\end{document}